# A Self-Healing Hardware Architecture for Safety-Critical Digital Embedded Devices

Shawkat S. Khairullah

Ph.D., Department of Computer Engineering, College of Engineering, University of Mosul, Mosul, Iraq

**ABSTRACT:** Digital Embedded Devices of next-generation safety-critical industrial automation systems require high levels of survivability and resilience against the hardware and software failure. One of the concepts for achieving this requirement is the design of resilient and survivable digital embedded systems. In the last two decades, development of self-healing digital systems based on molecular and cellular biology have received attention for the design of robust digital systems. However, many of these approaches have not been architected from the outset with safety in mind, nor have they been targeted for the applications of automation community where a significant need exists. This paper presents a new self-healing hardware architecture, inspired from the way nature responds, defends and heals: the stem cells in the immune system of living organisms, the life cycle of the living cell, and the pathway from Deoxyribonucleic acid (DNA) to protein. The proposed architecture is integrating cellular-based biological concepts, traditional fault tolerance techniques, and operational schematics for the international standard IEC 61131-3 to facilitate adoption in the automation industry and safety-critical applications. To date, two industrial applications have been mapped on the proposed architecture, which are capable of tolerating a significant number of faults that can stem from harsh environmental changes and external disturbances and we believe the nexus of its concepts can positively impact the next generation of critical systems in the automation industry.

## I. INTRODUCTION

Reliable digital embedded devices are rapidly being used in a variety of safety-related and safety-critical applications. Much digital electronic devices embedded in the aerospace, medical healthcare, automotive manufacturing, and nuclear industry have obvious consequences of failure. A failure of these safety-critical digital systems could result in death, serious injury to people, or severe damage to the environment [1], [6]. To date, many efforts have been made to design these digital devices using traditional redundancy and fault-tolerance techniques to achieve high levels of the safety requirements. A relatively new emerging research topics for the realization of resilient and robust digital embedded devices is the field of design biologically inspired self-healing digital system. It attempts to go beyond traditional fault tolerance techniques and redundancy approaches at the module level to learn from characteristics of nature and living organisms and adapting them to digital domain [1], [2], [3], [4]. We envision such bio-inspired self-healing reconfigurable machines being utilized in a wide variety of safety, life-critical applications, namely; cyber physical production systems, highly reliable advanced manufacturing, and safety-critical applications [5].

To achieve the promise of survivable behavior and resiliency in advanced safety-critical automation, these digital devices will need to rely on embedded controllers that have properties of resilience, fault-tolerance and self-healing. System resilience and self-healing computing methods are emergent technologies organized around a concept of self-governance – much in the way biological systems have evolved.

Among the many active areas of research in digital embedded devices research is reconfigurable computing (or processing), and resilient computing which is very active and enduring. Reconfigurable computing (RC) devices or units are systems or architectures (Hardware HW or Software SW) that are able to adapt to the application or environmental changes on the fly. For example, Field Programmable Gate Array (FPGA) contains reconfigurable logic blocks, memory elements, multiplexers, and routing units that can be modified (partially reconfigured) to perform the functionality of one application at one time and reconfigured again to match the new application at different time [7], [8], [9]. In contrast to General Purpose Processing, which executes a set of fixed instructions on fixed hardware sequentially, Reconfigurable Computing devices can change their internal hardware partially or completely at run-time by downloading a configuration bitstream file from a configuration memory while the system continues to function [10], [9]. Reconfigurable Computing is well suited for the emerging digital embedded devices needs to develop more resilient and self-healing architectures. Therefore, the need for new classes of reconfigurable systems that have the capability of being not only adaptable at run time, but also fault tolerant and self-healing in disruptive environmental conditions is highly needed.

## II. RELATED WORK ON SELF-HEALING DIGITAL SYSTEMS

The significance of building dependable computer-based digital systems has grown dramatically since von Newmann's seminal work on generating "reliable organisms from unreliable components" [11], [12]. The microelectronics industry is able to integrate silicon devices at nano-scales to synthesize complex commercial systems such as multi-core CPUs, System on a Chip (SoCs), and Graphical Processing Units (GPUs) as everyday commodities [10]. As "Moore's Law" begins to hit its limits, approaches to balance reliability, performance, and power consumption in digital Integrated Circuits (ICs) has become a significant research and development area. In [13], [14], [15], a new programmable cellular architecture that performs logic and arithmetic operations for a self-repairing FPGA has been designed by Mange et al. This architecture includes four hierarchal levels of organization: molecular, cellular, organismic, and population. At the molecular and cellular levels, two fault tolerant techniques based on time redundancy were used detect the fault occurrence inside the cell within the array at the decentralized level. In [16], [17], Tyrrell et al. have presented a different architecture which embeds a logic block performing the functions by a 2-1 multiplexer and a D flip-flop. This approach is a two-level hierarchical architecture consisting of a cellular level and organism level. Two modes, operation and configuration were proposed to control the operation of the organism in a fault tolerant manner. In [4], [18], [19], the researchers have been developed a biologically inspired, self-healing, resilient Instrumentation and Control (I&C) architecture, which is aimed to be used in next generation Nuclear Power Plant (NPP) digital applications that require high levels of fault-tolerance. The purpose of their work is to provide PLC (Programmable Logic Controller) like functionality with reliable constraints on execution behavior, and to maximize verifiability to address transient and permanent faults.

## III. OVERVIEW OF THE PROPOSED ARCHITECTURE

The proposed architecture is designed to achieve high levels of resilience, scalability, low complexity and reliability against different failure modes which are both hardware common cause failures and transient, intermittent, and permanent faults. The basic architecture shown in Fig. 1 (a) is based in part on the way biological living organisms achieve resiliency at the molecular and cellular levels. It is comprised of three principle divisions; (1) the Local Healing Layer that is corresponding to the operation of the life cycle of the cell; (2) the Critical Service Layer corresponding to B cells and T cells operation in the immune system, which is responsible for providing the intended functionality of the safety-critical application, and (3) the Global Healing Layer that is corresponding to embryonic stem cells operation, which is responsible for monitoring the behavior of the functions at the cellular level and triggering the required repair mechanisms to heal any faulty T cells present in the critical layer. We utilize these different layers to create interacting functional partitions to achieve overall system resilience (see Fig. 1 (b)).

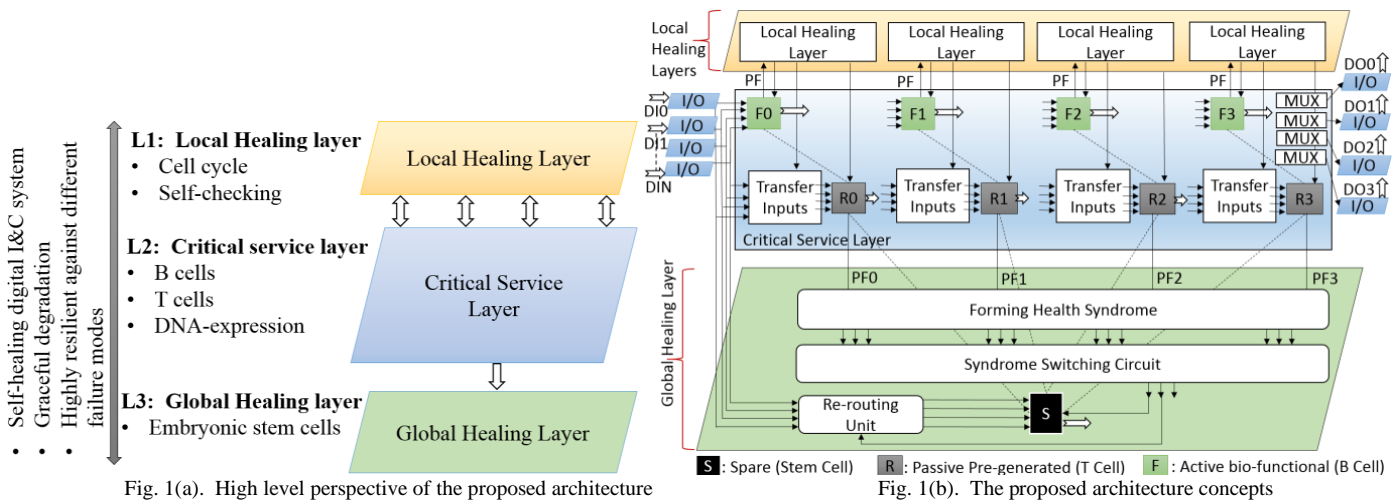

Fig. 1(a). High level perspective of the proposed architecture      Fig. 1(b). The proposed architecture concepts

Regarding the *critical service layer,* it executes the application functions. Specifically, it contains eight functional cells: *four* active B cells that are designated *F* in figure 1(b) and used to execute the application-based functions, and *four* passive pre-generated redundant T cells used as a healing mechanism for the faulty B cells designated *R* in figure1(b). The correct execution of each B cell is monitored continuously by its neighboring local healing layer, and once a transient fault is detected inside the input registers of each functional cell, it is tolerated using a hybrid redundancy unit embedded inside the same cell. The hybrid unit represents a first line of defense against the discovered transient faults defined as temporary deviations in the register values. The proposed architecture is designed to realize the functionality of safety-critical digital embedded systems operating in harsh environments, so radiation-induced transient faults that can occur at unpredictable times are the most prevalent fault type [20], [21], [22] and the process of correcting them at the functional block layer is most effective. If they are not tolerated at the block level, their wrong values will be sensitized at the output signals of the cell level, which is directly connected to the external world and can impact the safety of the public or the environment. On the other hand, once a permanent fault is detected in any one of the four B cells, its neighboring *local layer* is notified and generates a self-healing health syndrome, which executes three main tasks: (1) deactivates the output of the faulty B cell that simulates the operation of the cell death, (2) reroutes digital input data from the routing units of the faulty B cell and makes it available at the input of the routing units of the healthy T cell, which represents the reorganization process in the cell, and (3) selects a genetic code stored in a configuration memory in the T cell so that the functionality of the defected cell is healed and performed by the healthy T cell (restoration).

Finally, the *local healing layer* represents the second line of defense against the permanent faults effecting one or all of the four B cells. As a third line of defense against the occurrence of additional permanent faults, the *global healing layer* is responsible for fault management for the entire critical service layer.

IV. APPLICATIONS OF THE PROPOSED ARCHITECTURE

The bio-inspired self-healing digital system that has been designed and implemented is an evolving and active research project where refinements and new ideas are driven by mapping the architecture into different problem domains. To date, we have successfully mapped two application problems into the proposed architecture: Emergency Diesel Generator (EDG) Startup for Nuclear Power backup energy and an Automotive Cruise Control. These example applications are modest steps toward a planned at-scale-application related to complex distributed industrial control systems.

A. *Emergency Diesel Generator:*

The functional logic for the EDG, published in Electric Power Research Institute (EPRI) technical report is illustrated in Fig. 2. The EDG receives a total of fourteen digital input signals and produces two output signals. The output signals are calculated from the input signals using basic combinational logic AND, OR, and NOT operations. The EDG digital control system within a Nuclear Power Plant (NPP) is a safety critical system required for reactor cooling and other safety functionalities. While the functionality of the EDG is rather simplistic, it is a highly critical system that must be fault tolerant.

To demonstrate the resilience properties of the proposed architecture, the EDG critical application has been implemented on the architecture, which required fourteen functional cells to be connected together. Fig. 3(a), shows the simulation results for the first case study in which, three sequential transient faults have been injected into the three input registers of FAGFB for the first bio operational-cell (designated *F0* in figure 1(b)) in the architecture at times: 180ns, 240ns, and 300ns, hybrid redundancy units are tolerating their impacts quickly, and the output signal is generated without producing any erroneous value at the "Dat_Out_B" digital output port (see Fig. 3(a)). This signal represents the result of the GFB executing on the four digital input signals: "North", "West", "East", and "South".

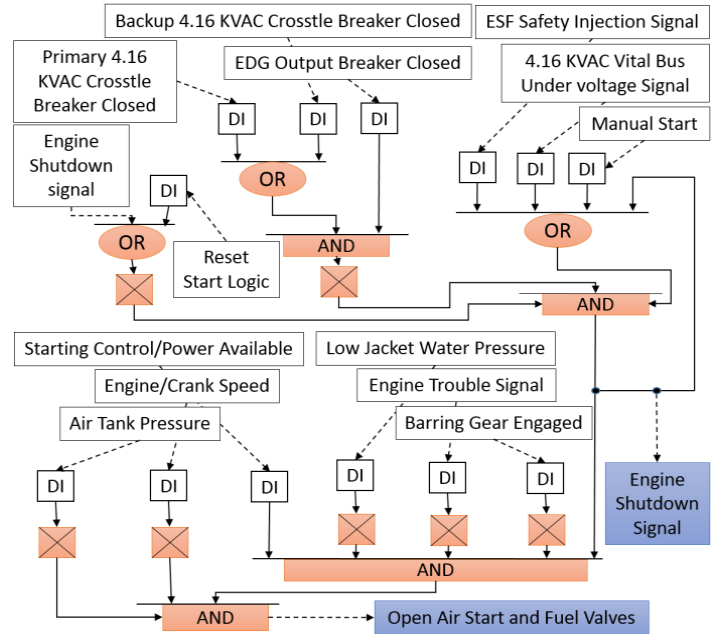

Fig. 2. Logic diagram for starting the EDG.

The second case study is shown in Fig. 3(b), in which, <u>two sequential permanent faults</u> have been injected into the GFB units of two cells: B cell (designated *F0* in figure 1(b)) and T cell (designated *R0* in figure 1(b)). In all cases, faults are identified by embedded self-checking units, self-healing is activated, and the system is repaired successfully. This type of multiple fault scenario typically occurs in digital I&C systems when there is cascading disturbance effect due to a power fluctuation, Electromagnetic Interference (EMI), and latch-up. Whenever a permanent fault is injected into the FAGFB of a cell, it will be detected immediately by a self-checking unit embedded in the same cell, and the three healing mechanisms start working in collaboration. For example, at time 850ns, as shown in Fig. 3(b), a permanent fault is injected into the FAGFB of the first B cell (designated *F0* in figure 1(b)) and the self-checking unit detects that fault. At this point, reconfiguration of the B-cell is initiated by transferring the four input signals of the faulty B cell to the inputs of the pre-generated T cell. Finally, the activation of the 66-bit genetic code stored in the pre-generated T cell completes the restoration process. The experimental results showed that the EDG application in a fault-free state requires an execution time of at least 245ns to produce the values of the two output ports: "EngineStart" and "OpenAirStartFuel_Valves". When the EDG is subjected to multiple permanent and transient faults into the GFB functional units of both B cell and T cell of the first critical-service layer, the EDG application is healed by time 570ns. This is about a 230 % increase in time delay to handle 4 faults – and this delay remains relatively constant as the number of handled faults increases.

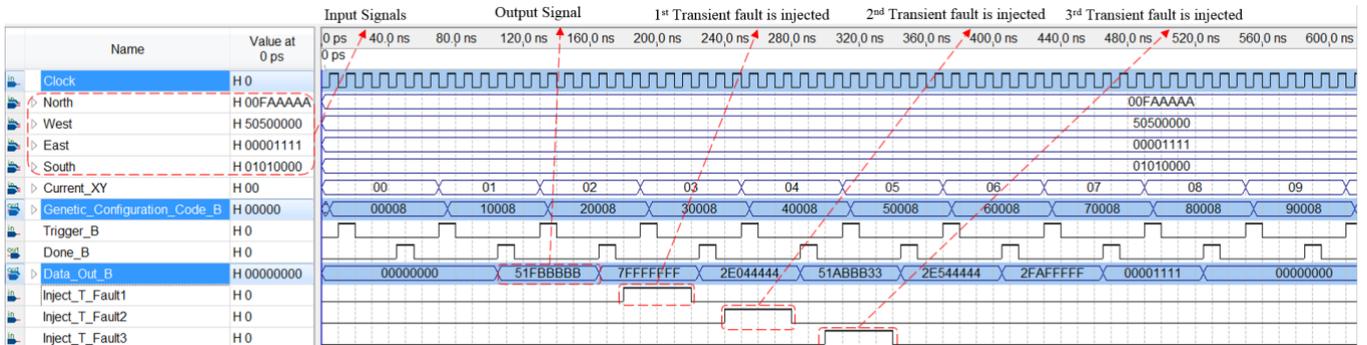

(a)

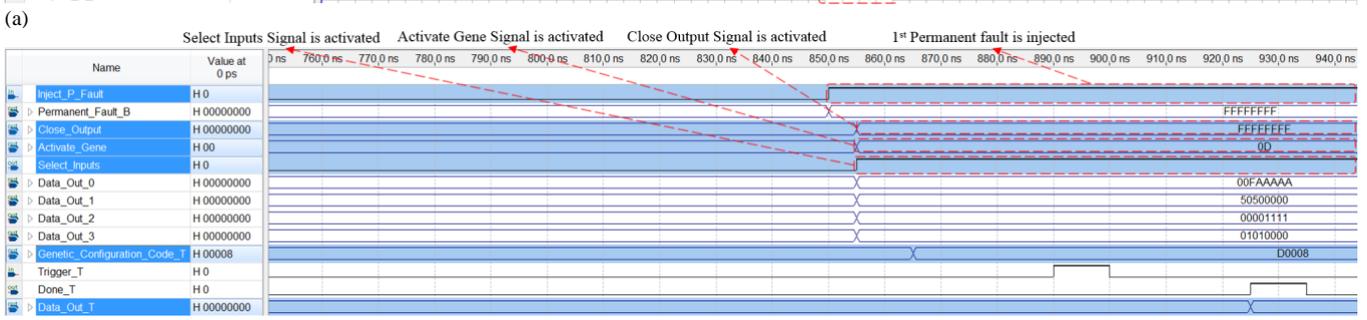

(b)

Fig. 3. (a) Time diagram simulation of injecting three sequential transient faults in the input registers of Bio-Operational B Cell. (b) Time diagram simulation of injecting one permanent fault in the 61131-Generic Functional Block.

B. *Cruise Control System:*

A classic example that illustrates mode-based control seen in process automation applications is the *automotive cruise control system (CCS)*. The CCS is a closed loop control system that keeps the vehicle tracking at a constant speed without depressing the accelerator pedal in spite of the external disturbances. This can be achieved by measuring the vehicle speed, comparing it to the desired speed, and then adjusting the throttle output value based on specific control rules like the Proportional Integral (PI) controller. The CCS receives a total of six digital input signals and produces two output signals. The output signals are calculated from the input signals using a combination of some digital control logic and a PI controller.

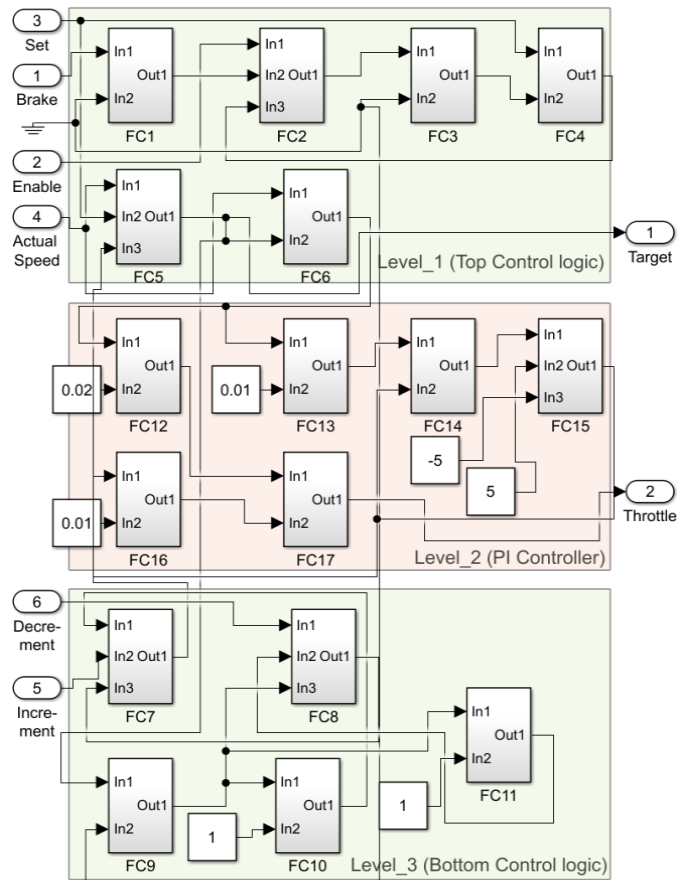

Fig. 4. A simplified block diagram for CCS implemented on the proposed architecture.

A block diagram for PI the controller that is used in many industrial control systems has been implemented on the proposed architecture with modest investment in time. To implement the CCS application on the architecture, this application has been partitioned into three levels: level1 (top control logic), level2 (PI controller), and level3 (bottom control logic) as shown in Fig. 4. Table I shows the different operations that are required to perform the mapping process of the CCS application and how they are distributed on 17 functional cells, and the four operational semantics of the designed CCS application are shown in Table II. As a consequence, the functionalities of these three different levels were distributed among five critical-service layers of the architecture illustrated in Fig. 1(b), and at each layer four functional cells (designated *F*) are triggered at a specified time. These five layers have been connected in a distributed way (see Fig. 4) to execute two tasks: Task1 represents the top and bottom control logic and Task2 represents the PI controller in such a way that Task1 requires three critical service layers and Task2 requires only two critical service layers.

TABLE I
CCS FUNCTIONALITY IS MAPPING ON 17 FUNCTIONAL CELLS

| Implementation Level | Functional Cell | Operation |
|---|---|---|
| Level_1 | FC1 | NOT |
| Top | FC2 | Addition |
| Control logic | FC3 | Delay |
| | FC4 | OR |
| | FC5 | Multiplexing |
| | FC6 | Subtraction |
| Level_2 | FC12, FC13 | Multiplication |
| PI Controller | FC14, FC17 | Addition |
| | FC15 | Comparison |
| | FC16 | Multiplication |
| Level_3 | FC7, FC8 | Multiplexing |
| Bottom | FC9 | Delay |
| Control logic | FC10 | Addition |
| | FC11 | Subtraction |

TABLE II
OPERATIONAL SEMANTICS OF CCS APPLICATION

| Condition | State | Operation |
|---|---|---|
| Set | Speed is set | Target speed = Actual speed |
| Decrement | Speed is decreased | Target speed = target speed -1 |
| Increment | Speed is increased | Target speed = target speed +1 |
| Cancel/Brake | Speed is cancelled | Target speed = 0 |

## V. Conclusion and Future Work

A new design and application of a biologically inspired self-healing hardware architecture has been presented in this paper. The architecture is designed to be utilized in safety-critical applications where; robustness against different failure modes is highly important, and flexibility to change the actual hardware platform and reconfigure the faulty H/W components is essential. Run-time simulation results and traditional fault injection methods indicate that the proposed self-repairing multi-layered approach to resilience strikes a good balance between the local fault detection layers and the global fault decisions layers that are represented by global healing layer to maintain continuity of system functionality and resilience. One of the interesting future works for this project is that the design of the self-healing architecture will be automated and the process of adding the repairing logic manually will be replaced. Finally, FPGA/ASIC design and verification automation methodologies and tools will be investigated to come up with standard format.